\documentclass[aps,pre,floatfix,twocolumn,epsfig]{revtex4}

\usepackage{amssymb,epsf}
\usepackage{graphicx} 
\usepackage{dcolumn} 
\usepackage{bm}
\usepackage{epsfig}

\begin{document}

\title{The architecture of complex weighted networks}

\author{Alain Barrat$^1$, Marc Barth{\'e}l\'emy$^2$,
Romualdo Pastor-Satorras$^3$, and
Alessandro Vespignani$^1$}

\affiliation{$^1$ Laboratoire de Physique Th{\'e}orique (UMR du CNRS 8627),
  Batiment 210, Universit{\'e} de Paris-Sud 91405 Orsay, France\\
$^2$CEA-Centre d'Etudes de
  Bruy{\`e}res-le-Ch{\^a}tel, D\'epartement de Physique Th\'eorique et
  Appliqu\'ee BP12, 91680 Bruy{\`e}res-Le-Ch{\^a}tel, France\\
$^3$Departament de F\'isica i
  Enginyeria Nuclear Universitat Polit{\`e}cnica de Catalunya, Campus Nord
  08034 Barcelona, Spain}

\date{\today}

\begin{abstract}
Networked structures arise in a wide array of different contexts such as
technological and transportation infrastructures, social phenomena,
and biological systems.  These highly interconnected systems have
recently been the focus of a great deal of attention that has
uncovered and characterized their topological complexity.  Along with
a complex topological structure, real networks display a large
heterogeneity in the capacity and intensity of the connections. These
features, however, have mainly not been considered in past studies
where links are usually represented as binary states, i.e. either
present or absent. Here, we study the scientific collaboration network
and the world-wide air-transportation network, which are
representative examples of social and large infrastructure systems,
respectively.  In both cases it is possible to assign to each edge of
the graph a weight proportional to the intensity or capacity of the
connections among the various elements of the network.  
We define new appropriate  metrics combining weighted and topological
observables that enable us to characterize the complex statistical 
properties and heterogeneity of the actual strength of edges and 
vertices. This information allows us to investigate for the first 
time the correlations among weighted quantities and the underlying 
topological structure of the network. These results provide a 
better description of the hierarchies and organizational principles 
at the basis of the architecture of weighted networks.
\end{abstract}

\maketitle 

\section{Introduction}
A large number of natural and man-made systems are structured in the
form of networks. Typical examples include large communication systems
(the Internet, the telephone network, the World-Wide-Web),
transportation infrastructures (railroad and airline routes),
biological systems (gene and/or protein interaction networks), and a
variety of social interaction structures.
\cite{barabasi99,barabasi02,mendesbook,amaral}. The macroscopic properties of
these networks have been the subject of an intense scientific activity
that has highlighted the emergence of a number of significant
topological features. Specifically, many of these networks show the
small-world property \cite{watts98}, which implies that the network
has an average topological distance between the various nodes
increasing very slowly with the number of nodes (logarithmically or
even slower), despite showing a large degree of local
interconnectedness typical of more ordered lattices. Additionally,
several of these networks are characterized by a statistical abundance
of ``hubs'' with a very large number of connections $k$ compared to
the average degree value $\langle k \rangle$.  The empirical evidence
collected from real data indicates that this distinctive feature finds
its statistical characterization in the presence of scale-free degree
distributions $P(k)$, i.e., showing a power-law behavior $P(k) \sim
k^{-\gamma}$ for a significant range of values of $k$
\cite{barabasi02,mendesbook}.  These topological features turn out to
be extremely relevant since they have a strong impact in assessing
such networks' physical properties as their robustness or
vulnerability \cite{havlin00,newman00,barabasi00,pv01a}.

While these findings alone might provide possibilities for threats
analysis and policy decisions, yet networks are not only specified by
their topology but also by the dynamics of information or traffic flow
taking place on the structure.  In particular, the heterogeneity in
the intensity of connections may be very important in the
understanding of social systems. Analogously, the amount of traffic
characterizing the connections of communication systems or the
quantity of traffic in large transport infrastructures is fundamental
for a full description of these networks.

Motivated by these observations, we undertake in this paper the
statistical analysis of complex networks whose edges have assigned a
given weight (the flow or the intensity) and thus can be generally
described in terms of \textit{weighted graphs}
\cite{weightbook,yookwt}.  Working with two typical examples of this
kind of networks, we introduce some new metrics that combine in a
natural way both the topology of the connections and the weight
assigned to them. These quantities provide a general characterization
of the heterogenous statistical properties of weights and identify
alternative definitions of centrality, local cohesiveness, and
affinity.  By appropriate measurements it is also possible to exploit
the correlation between the weights and the topological structure of
the graph, unveiling the complex architecture shown by real weighted
networks.\\

\section{Weighted networks data}

In order to proceed to the general analysis of complex weighted networks we
consider two specific examples for which it is possible to have a full
characterization of the links among the elements of the systems, the 
world-wide airport transportation network  and  the
scientist collaboration network.

{\em World-wide Airport Network (WAN)}: We analyze the International
Air Transportation Association (IATA) \footnote{http://www.iata.org.}
database containing the world list of airports pairs connected by
direct flights and the number of available seats on any given
connection for the year 2002. The resulting air-transportation graph
comprises $N=3880$ vertices denoting airports and $E=18810$ edges
accounting for the presence of a direct flight connection.  The
average degree of the network is $\langle k \rangle =2E/N=9.70$, while the
maximal degree is $318$.  The topology of the graph exhibits both
small-world and scale-free properties as already observed in different
dataset analyses \cite{china,luisair}. In particular, the average
shortest path length, measured as the average number of edges
separating any two nodes in the network, shows the value
$\langle\ell\rangle=4.37$, very small compared to the network size
$N$. The degree
distribution, on the other hand, takes the form $P(k)=
k^{-\gamma}f(k/k_x)$, where $\gamma\simeq2.0$ and $f(k/k_x)$ is an exponential
cut-off function, that finds its origin in physical constraints on the
maximum number of connections that a single airport can handle
\cite{amaral,luisair}. The airport connection graph is therefore a clear
example of heterogeneous network showing scale-free properties on a
definite range of degree values.

{\em Scientific Collaboration Network (SCN)}: We consider the network
of scientists who have authored manuscripts submitted to the e-print
archive relative to condensed matter physics
(http://xxx.lanl.gov/archive/cond-mat) between 1995 and
1998. Scientists are identified with nodes and an edge exists between
two scientists if they have co-authored at least one paper. The
resulting connected network has $N=12722$ nodes, with an average
degree (i.e. average number of collaborators) $\langle k \rangle=6.28$
and maximal degree $97$.  The
topological properties of this network and other similar networks of
scientific collaborations have been studied in Ref.~\cite{newman01,vicsek01}.

\begin{figure}
\centerline{
\includegraphics*[width=0.45\textwidth]{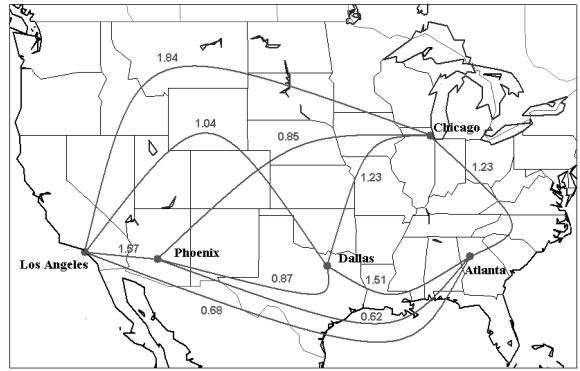}
}
\caption{Pictorial representation of the weighted graph obtained 
from the  airport network
data. Major US airports are connected by edges denoting the presence
of a non-stop flight in both directions whose weights represent the
number of avaliable seats (million/year).}
\label{fig0}
\end{figure}

The properties of a graph can be expressed via 
its adjacency matrix $a_{ij}$, whose elements take the value $1$ if an
edge connects the vertex $i$ to the vertex $j$ and $0$ otherwise. The
data contained in the previous datasets permit to go beyond this topological
representation by defining a weighted graph \cite{weightbook} that
assigns a weight or value characterizing each connecting link. In the
case of the WAN the weight $w_{ij}$ of an edge linking airports $i$
and $j$ represents the number of available seats in flights between
these two airports.  The inspection of the weights shows that the 
average numbers of seats in both directions are 
identical $w_{ij}=w_{ji}$ for an overwhelming
majority of edges. In the following we will thus work with the
symmetric undirected graph and avoid the complication deriving from
flow imbalances. We show an example of the resulting weighted 
graph in Fig.~\ref{fig0}. Noticeably, the above definition of weights 
is a straightforward and an objective measure of the traffic flow 
on top of the network.

For the SCN we follow the definition of weight introduced in
Ref.~\cite{newman01}: The intensity $w_{ij}$ of the interaction 
between two collaborators $i$ and $j$ is defined as
\begin{equation}
  w_{ij} = \sum_p \ \frac{\delta_i^p \delta_j^p}{n_p -1}\ ,
\end{equation}
where the index $p$ runs over all papers, $n_p$ is the number of
authors of the paper $p$, and $\delta_i^p$ is $1$ if author $i$ has
contributed to paper $p$, and $0$ otherwise.  While any definition of
the intensity of a connection in social networks is depending on the
particular elements chosen to be relevant, the above definition seems
to be rather objective and representative of the scientific
interaction: It is large for collaborators having many papers in
common but the contribution to the weight introduced by any given
paper is inversely proportional to the number of
authors.\\

\section{Centrality and weights}

In order to take into account the information provided by the weighted
graph, we shall identify the appropriate quantities characterizing it
structure and organization at the statistical level. The statistical
analysis of weights $w_{ij}$ between pairs of vertices indicates the
presence of right skewed distributions, already signaling a high level
of heterogeneity in the system for both the WAN and the SCN as also
reported in Refs.~\cite{china,newman01}.  It has been observed,
however, that the individual edge weights do not provide a general
picture of the network's complexity~\cite{yookwt}.  A more
significative measure of the network properties in terms of the actual
weights is obtained by looking at the vertex {\em strength} $s_i$,
defined as
\begin{equation}
  s_i=\sum_{j=1}^N a_{ij}w_{ij}\,.
\label{Eq_strength}
\end{equation}
This quantity measures the strength of vertices in terms of the total
weight of their connections.  In the case of the WAN the vertex
strength simply accounts for the total traffic handled by each
airport. For the SCN, on the other hand, the strength is a measure of
scientific productivity since it is equal to the total number of
publications of any given scientist. This quantity is a natural
measure of the importance or centrality of a vertex $i$ in the
network.

The identification of the most central nodes in the system is a major
issue in networks characterization~\cite{freeman77}.  The most
intuitive topological measure of centrality is given by the
degree---more connected nodes are more central.  However, more is not
necessarily better.  Indeed, by considering solely the degree of a
node we overlook that nodes with small degree may be crucial for
connecting different regions of the network by acting as bridges.  In
order to quantitatively account for the role of such nodes it has been
introduced the concept of betweenness centrality
\cite{freeman77,newman01,goh01}, that is defined as the number of shortest
paths between pairs of vertices that pass through a given
vertex~\cite{notebet}. Central nodes are therefore part of more
shortest paths within the network than peripheral nodes. Moreover, the
betweenness centrality is often used in transport networks to provide
an estimate of the traffic handled by the vertices, assuming that the
number of shortest paths is a zero-th order approximation to the
frequency of use of a given node \footnote{For the airport network,
  the analysis of the betweenness centrality and its correlation with
  the degree has been discussed in Ref.~\cite{luisair}.}.  The above
definition of centrality relies only on topological elements. It is
therefore intuitive to consider the alternative definition of
centrality constructed by looking at the strength $s_i$ of the
vertices as a more appropriate definition of the importance of a
vertex in weighted networks. For instance, in the case of the WAN this
quantity provides the actual traffic going through the vertex $i$, and
it is natural to study how it compares and correlates with the
topological measures of centrality.

\begin{figure}
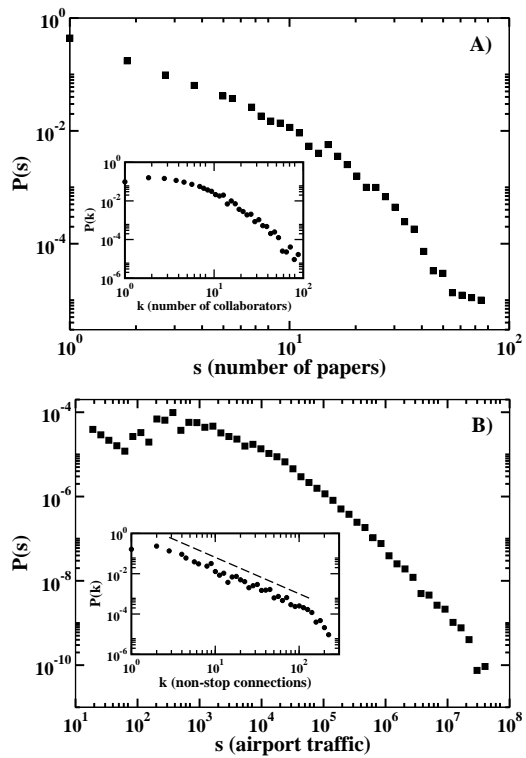

\centerline{
\includegraphics*[width=0.38\textwidth]{fig1a}
}
\vspace*{0.2cm}
\centerline{
\includegraphics*[width=0.38\textwidth]{fig1b}
}
\caption{ {\bf A} Degree and strength distribution 
  in the scientific collaboration network. The degree $k$ corresponds
  to the number of co-authors of each scientist and the strength
  represent its total number of publications. The distributions are
  heavy-tailed even if it is not possible to distinguish a definite
  functional form.
  {\bf B} The same distributions for the world-wide airport network.
  The degree is the number of non-stop connections to other airports
  and the strength is the total number of passengers handled by any
  given airport. In this case, the degree distribution can be
  approximated by the power-law behavior $P(k)\sim k ^{-\gamma}$ with
  $\gamma=1.8\pm 0.2$. The strength distribution has a heavy-tail extending
  over more than four orders of magnitude.}
\label{fig1}
\end{figure}

\begin{figure}
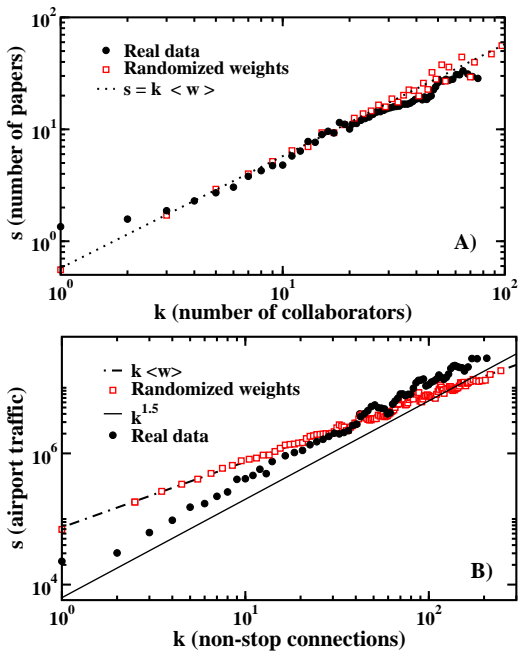

\centerline{
\includegraphics*[width=0.38\textwidth]{fig2a}
}
\vspace*{.1cm}
\centerline{
\includegraphics*[width=0.38\textwidth]{fig2b}
}
\caption{  Average strength $s(k)$ as function of the degree $k$ of nodes. 
  {\bf A} In the scientific collaboration network the real data are
  very similar to those obtained in a randomized weighted network.
  Only at very large $k$ values it is possible to observe a slight
  departure from the expected linear behavior.  {\bf B} In the world
  airport network real data follow a power-law behavior with exponent
  $\beta=1.5\pm0.1$. This denotes anomalous correlations between the
  traffic handled by an airport and the number of its connections.}
\label{fig2}
\end{figure}

The probability distribution $P(s)$ that a vertex has strength $s$ is
heavy tailed in both networks and the functional behavior exhibits
similarities with the degree distribution $P(k)$ (see
Fig.~\ref{fig1}). A precise functional description of the heavy-tailed
distributions may be very important in understanding the network
evolution and will be deferred to future analysis. This behavior is
not unexpected since it is plausible that the strength $s_i$ increases
with the vertex degree $k_i$, and thus the slow decaying tail of
$P(s)$ stems directly from the very slow decay of the degree
distribution.
%
%
In order to shed more light on the relation between the vertices'
strength and degree, we investigate the dependence of $s_i$ on $k_i$.
We find that the average strength $s(k)$ of vertices with degree $k$
increases with the degree as
\begin{equation}
  s(k) \sim k^{\beta} .
\end{equation}
In the absence of correlations between the weight of edges and the
degree of vertices, Eq.~(\ref{Eq_strength}) implies that $s(k)= \langle w
\rangle k$, where $\langle w \rangle =(2E)^{-1}\sum_{i,j} a_{ij} w_{ij}$ is the average
weight in the network. The strength of a vertex is then simply
proportional to its degree, yielding an exponent $\beta=1$, and the two
quantities provide therefore the same information on the system. In
Fig.~\ref{fig2} we report the behavior obtained for both the real
weighted networks and their randomized versions, generated by a random
re-distribution of the actual weights on the existing topology of the
network.  For the SCN the curves are very similar and well fitted by
the uncorrelated approximation $s(k)= \langle w \rangle k$.  Strikingly, this is
not the case of the WAN.  Fig.~\ref{fig2}B clearly shows a very
different behavior for the real data set and its randomized version.
In particular, the power-law fit for the real data gives an
``anomalous'' exponent $\beta_{\rm WAN}=1.5\pm 0.1$.  This implies that the
strength of vertices grows faster than their degree, i.e.  the weight
of edges belonging to highly connected vertices tends to have a value
higher than the one corresponding to a random assignment of weights.
This denotes a strong correlation between the weight and the
topological properties in the WAN, where the larger is an airport, the
more traffic it can handle.

\begin{figure} 
\vspace*{.1cm} \centerline{
\includegraphics*[width=0.38\textwidth]{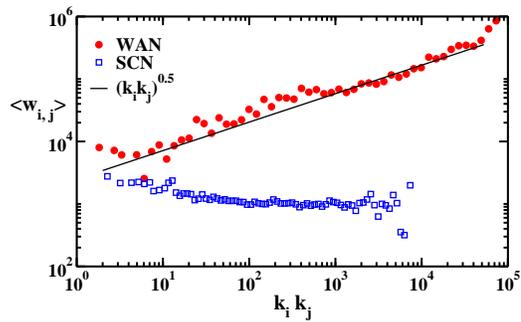} }
\caption{Average weight as a function of the degree end points. The
  full line corresponds to a power-law behavior $\langle w_{i,j} \rangle \sim (k_i
  k_j)^\theta$, with exponent $\theta=0.5\pm0.1$. In the case of the scientific
  collaboration network it is possible to observe an almost flat
  behavior for roughly three orders of magnitude.}
\label{fig3}
\end{figure}

The fingerprint of these correlations is also observed in the
dependence of the weight $w_{i j}$ on the degrees of the end point
nodes $k_i$ and $k_j$. As we can see in Fig.~\ref{fig3}, for the WAN
the behavior of the average weight as a function of the end points
degrees can be well approximated by a power-law dependence
\begin{equation}
  \langle w_{ij} \rangle \sim(k_ik_j)^{\theta}\,
\end{equation}
with an exponent $\theta=0.5\pm 0.1$.  This exponent can be related to
the $\beta$ exponent by noticing that $s(k) \sim k(kk_j)^{\theta}$,
resulting in $\beta=1+\theta$, if the topological correlations
between the degree of connected vertices can be neglected. This is
indeed the case of the WAN where the above  scaling relation 
is well satisfied by the numerical values provided by the independent 
measurements of the exponents. 
In the SCN, instead, $\langle w_{ij} \rangle$ is almost constant for over
two decades confirming a general lack of correlations between 
the weights and the vertices degree.

Analogously, a study of the average value $s(b)$ of the strength
for vertices with betweenness $b$ shows that the functional
behavior can be approximated by a scaling form
$s(b) \sim b^{\delta}$ with $\delta_{\rm SCN}\simeq 0.5$ 
and $\delta_{\rm WAN}\simeq0.8$ for the
SCN and the WAN, respectively. As before, the comparison between the
behavior of the real data and the randomized case shows more
pronounced differences in the case of the WAN.
In both networks, the strength grows with the betweenness faster than
in the randomized case, specially in the WAN. This is
another clear signature of the correlations between weighted
properties and the network topology.

\section{Structural organization of weighted networks}

Along with the general vertices hierarchy imposed by the strength
distribution---the larger the more central---complex networks show an
architecture imposed by the structural and administrative organization
of these systems. For instance, topical areas and national research
structures give rise to well defined groups or communities in the SCN.
In the WAN, on the other hand, different hierarchies correspond to
domestic or regional airport groups and intra-continental transport
systems; political or economical factors can as well impose additional
constraints to the network structure~\cite{luisair}.  In order to
uncover these structures, some topological quantities are customarily
studied.  The clustering coefficient measures the local group
cohesiveness and is defined for any vertex $i$ as the fraction of
connected neighbors of $i$ \cite{watts98}. The average clustering
coefficient $C=N^{-1}\sum_i c_i$ thus expresses the statistical level of
cohesiveness measuring the global density of interconnected vertices'
triples in the network.  Further information can be gathered by
inspecting the average clustering coefficient $C(k)$ restricted to
classes of vertices with degree $k$. Often, $C(k)$ exhibits a highly
non-trivial behavior with a power-law decay as a function of $k$,
signaling a hierarchy in which low degree vertices belong generally to
well interconnected communities (high clustering coefficient), while
hubs connect many vertices that are not directly connected (small
clustering coefficient) \cite{alexei02,ravasz02}.  Another quantity
used to probe the networks' architecture is
the behavior of the average degree of nearest neighbors, $k_{nn}(k)$,
for vertices of degree $k$ \cite{alexei}.  This last quantity is
related to the correlations between the degree of connected vertices
\cite{maslov02}
since it can be expressed as $k_{nn}(k)=\sum_{k'} k'P(k'|k)$, where
$P(k'|k)$ is the conditional probability that a given vertex with
degree $k$ is connected to a vertex of degree $k'$. In the absence of
degree correlations, $P(k'|k)$ does not depend on $k$ and neither does
the average nearest neighbors' degree; i.e. $k_{nn}(k)= {\rm const}.$
\cite{alexei}. In the presence of correlations, the behavior of
$k_{nn}(k)$ identifies two general classes of networks. If $k_{nn}(k)$
is an increasing function of $k$, vertices with high degree have a
larger probability to be connected with large degree vertices. This
property is referred in physics and social sciences as {\em assortative
  mixing} \cite{assortative}. On the contrary, a decreasing behavior
of $k_{nn}(k)$ defines {\em disassortative mixing}, in the sense that
high degree vertices have a majority of neighbors with low degree,
while the opposite holds for low degree vertices.

\begin{figure} 
\vspace*{.1cm} \centerline{
\includegraphics*[width=0.42\textwidth]{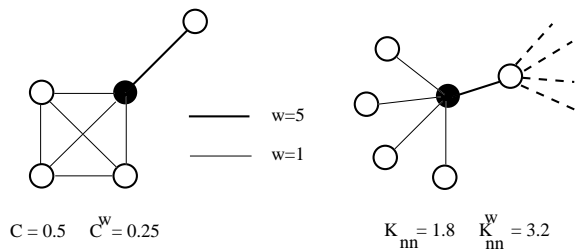} }
\caption{Examples of local configurations whose topological and
weighted quantities are different. In both cases the central vertex
(filled) has a very strong link with only one of its neighbors. 
The weighted clustering and average nearest neighbors degree
capture more precisely the effective level of cohesiveness and
affinity due to the actual interaction strength.}
\label{figdis}
\end{figure}

The above quantities provide clear signatures of a structural
organization of networks in which different degree classes show
different properties in the local connectivity structure.  However,
they are defined solely on topological grounds and the inclusion of
weights and their correlations might change consistently our view of
the hierarchical and structural organization of the network 
(see Fig.~\ref{figdis}).  This can be easily understood with the 
simple example of a network in which the
weights of all edges forming triples of interconnected vertices are
extremely small. Even for a large clustering coefficient, it is clear
that these triples have a minor role in the network dynamics and
organization, and that the clustering properties are definitely
overestimated by a simple topological analysis. Similarly, high degree
vertices could be connected to a majority of low degree vertices while
concentrating the largest fraction of their strength only on the
vertices with high degree. In this case the topological information
would point out to disassortative properties while the network could
be considered assortative in an effective way, since the more
relevant edges in term of weights are linking high degree vertices.

In order to solve the previous incongruities we introduce new metrics
that combine the topological information with the weight distribution
of the network.  First, we consider the {\em weighted clustering
coefficient} defined as
\begin{equation}
  c_{i}^w=\frac{1}{s_i(k_i-1)} \sum_{j, h}
  \frac{(w_{ij}+w_{ih})}{2} a_{ij}a_{ih}a_{jh}.
\end{equation}
This is a measure of the local
cohesiveness that takes into  account the importance of the 
clustered structure on the basis of the amount of traffic or
interaction intensity actually found on the local triples. 
Indeed, $c_{i}^w$ is counting for each triple 
formed in the neighborhood of the vertex $i$ the weight of 
the two participating edges of the vertex $i$. In this way we are  
not just considering the number of closed triangles
in the neighborhood of a vertex but also their total relative weight 
with respect to the vertex' strength. The normalization factor 
$s_i(k_i-1)$ account for the weight of each edge times the maximum 
possible number of triangles it may participate, and it ensures  that 
$0\leq c_{i}^w\leq 1$. Consistently, the  $c_{i}^w$ definition
recovers the topological clustering coefficient in the case that 
$w_{ij}=const$. 
Next we define $C^w$ and $C^w(k)$ as the weighted clustering
coefficient averaged over all vertices of the network and over all
vertices with degree $k$, respectively. This quantities provide global
information on the correlation between weights and topology, specially
by comparing them with their topological analogs.
In the case of a large randomized
network (lack of correlations) it is easy to see that $C^w=C$ and
$C^w(k)=C(k)$. In real weighted networks,
however, we can face two opposite cases. If $C^w > C$, we are in
presence of a network in which the interconnected triples are more
likely formed by the edges with larger weights. On the contrary,
$C^w<C$ signals a network in which the topological clustering is
generated by edges with low weight. In this case it is obvious that
the clustering has a minor effect in the organization of the network
since the largest part of the interactions (traffic, frequency of the
relations, etc.) is occurring on edges not belonging to interconnected
triples. The same may happen for $C^w(k)$, for which it is also
possible to analyze the variations with respect to the degree class
$k$.
%
\begin{figure}
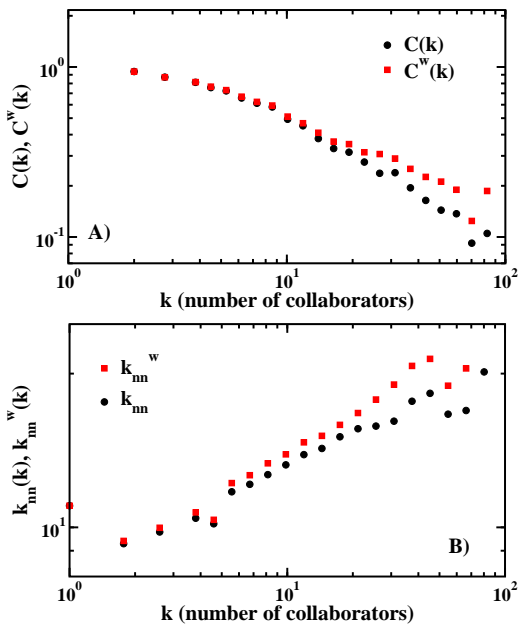
 
\centerline{
\includegraphics*[width=0.38\textwidth]{fig5a}
}
\vspace*{.1cm}
\centerline{
\includegraphics*[width=0.38\textwidth]{fig5b}
}
\caption{ Topological and weighted quantities for the SCN. 
  {\bf A} The weighted clustering separates form  the topological
  one around $k\geq 10$. This marks a difference for authors with 
  larger number of collaborators.
  {\bf B} The assortative behavior is enhanced in the weighted
  definition of the average nearest neighbors degree.}
\label{fig5}
\end{figure}

\begin{figure}
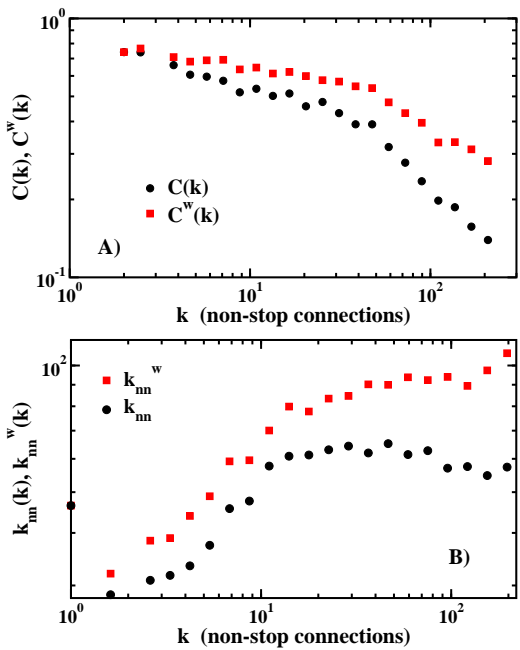
 
\centerline{
\includegraphics*[width=0.38\textwidth]{fig6a}
}
\vspace*{.1cm}
\centerline{
\includegraphics*[width=0.38\textwidth]{fig6b}
}
\caption{ Topological and weighted quantities for the WAN.
  {\bf A} The weighted clustering coefficient is larger than 
  the topological one in the whole degree spectrum.
  {\bf B} $k_{nn}(k)$ is reaching a plateau for  $k>10$ 
  denoting the absence of marked topological correlations.
  On the contrary $k^w_{nn}(k)$ exhibits  a more definite 
  assortative behavior.}
\label{fig6}
\end{figure}
%

Along with the weighted clustering coefficient, we introduce the {\em
weighted average nearest neighbors degree}, defined as
\begin{equation}
  k^w_{nn,i}=\frac{1}{s_i}\sum_{j=1}^N a_{ij} w_{ij} k_j.
\end{equation}
In this case, we perform a local weighted average of the nearest
neighbor degree according to the normalized weight of the connecting
edges, $w_{ij} / s_i$.  This definition implies that $k^w_{nn,i}>
k_{nn,i}$ if the edges with the larger weights are pointing to the
neighbors with larger degree and $k^w_{nn,i}< k_{nn,i}$ in the
opposite case. The $k^w_{nn,i}$ thus measures the effective {\em
  affinity} to connect with high or low degree neighbors according to
the magnitude of the actual interactions.  As well, the behavior of
the function $k^w_{nn}(k)$ marks the weighted assortative or
disassortative properties considering the actual interactions among
the system's elements.

As a general test, we inspect the results obtained for both the SCN
and the WAN by comparing the regular topological quantities with those
obtained with the weighted definition introduced here. The topological
measurements tell us that the SCN has a continuously decaying spectrum
$C(k)$ (see Fig.~\ref{fig5}a). This implies that hubs present a much
lower clustered neighborhood than low degree vertices.  This effect
can be interpreted as the evidence that authors with few collaborators
usually work within a well defined research group in which all the
scientists collaborate together (high clustering).  Authors with a
large degree, however, collaborate with different groups and
communities which on their turn do not have often collaborations, thus
creating a lower clustering coefficient.  Furthermore, the SCN
exhibits an assortative behavior in agreement with the general
evidence that social networks are usually denoted by a strong
assortative character \cite{assortative} (see Fig.~\ref{fig5}b).  The
analysis of weighted quantities confirms this topological picture,
providing further information on the network architecture.  The 
weighted clustering coefficient is very close to the topological
one ($C^w/C\simeq 1$).  This fact states in a quantitative way that group
collaborations tend on average to be stable and determine the average
intensity of the interactions in the network. In addition, the
inspection of $C^w(k)$ (see Fig.~\ref{fig5}a) shows generally 
that for $k\geq 10$ the weighted clustering coefficient is larger than
the topological one. This implies that high
degree authors (i.e. with many collaborators) tend to publish
more papers with interconnected groups of  co-authors. This is
eventually a signature that influential scientists form stable
research groups where the largest part of their production is obtained.
Finally, the 
assortative properties find a clearcut confirmation in the weighted 
analysis with a $k^w_{nn}(k)$ strikingly growing as a power-law as a function of
$k$.

A different picture is found in the WAN, where the weighted analysis
provides a richer and somehow different scenario.  This network also
shows a decaying $C(k)$, consequence of the role of large airports
that provide non-stop connections to very far destinations on an
international and intercontinental scale. These destinations are
usually not interconnected among them, giving rise to a low clustering
coefficient for the hubs. We find, however, that $C^w/C\simeq 1.1$,
indicating an accumulation of traffic on interconnected groups of
vertices. The weighted clustering coefficient $C^w(k)$ also has a
different behavior in that its variation is much more limited in the
whole spectrum of $k$. This implies that high degree airports have a
progressive tendency to form interconnected groups with high traffic
links, thus balancing the reduced topological clustering. Since high
traffic is associated to hubs, we have a network in which high degree
nodes tend to form cliques with nodes with equal or higher degree, the
so-called {\em rich-club phenomenon} \cite{richclub}.  An interesting
evidence emerges also from the comparison of $k_{nn}(k)$ and
$k^w_{nn}(k)$. Indeed, the topological $k_{nn}(k)$ does show an
assortative behavior only at small degrees.  For $k>10$, $k_{nn}(k)$
approaches a constant value, a fact revealing an uncorrelated
structure in which vertices with very different degrees have a very
similar neighborhood. The analysis of the weighted $k^w_{nn}(k)$,
however, exhibits a pronounced assortative behavior in the whole $k$
spectrum, providing a different picture in which high degree airports
have a larger affinity for other
large airports where the major part of the traffic is directed.

\section{Conclusions}

We have shown that a more complete view of complex networks is
provided by the study of the interactions defining the links of these
systems. The weights characterizing the various connections exhibit
complex statistical features with highly varying distributions and
power-law behavior.  In particular we have considered the specific
examples of the scientific collaboration and world-wide airport
networks where it is possible to appreciate the importance of the
correlations between weights and topology in the characterization of
real networks properties. Indeed, the analysis of the weighted
quantities and the study of the correlations between weights and
topology provide a complementary perspective on the structural
organization of the network that might be  undetected by quantities
based only on topological information. Our study  thus 
offer a quantitative and general approach to understand 
the complex architecture of real weighted networks.\\

\begin{acknowledgments}
{\small We thank IATA for making the airline commercial flight
  database available to us. We also thank M.E.J. Newman for conceding
  us the possibility of using the scientific collaboration network
  data (see http://www-personal.umich.edu/$\sim$mejn/collaboration/).
  We are grateful to L. A. N. Amaral and R. Guimer{\`a} for the many
  discussions and sharing of results during the various stages of this
  work.  A.B., R.P.-S. and A.V. are partially funded by the 
  European Commission
  - Fet Open Project COSIN IST-2001-33555. R.P.-S.  acknowledges
  financial support from the Ministerio de Ciencia y Tecnolog{\'\i}a
  (Spain) and from the Departament d'Universitats, Recerca i Societat
  de la Informaci{\'o}, Generalitat de Catalunya (Spain).}
\end{acknowledgments}


\begin{thebibliography}{10}

\bibitem{barabasi99}
 Barab{\'a}si, A.-L \& Albert, R. 
\newblock (1999) {\em Science} {\bf 286}, 509.

\bibitem{barabasi02}
Albert, R. \& Barab{\'a}si, A.-L.
\newblock (2002) {\em Rev. Mod. Phys.} {\bf 74}, 47--97.

\bibitem{mendesbook}
Dorogovtsev, S.~N. \& Mendes, J. F.~F.
\newblock (2003) {\em Evolution of networks: From biological nets to the
  {I}nternet and {WWW}}.
\newblock (Oxford University Press, Oxford).

\bibitem{amaral}
Amaral, L. A.~N., Scala, A., Barth{\'e}lemy, M,  \& Stanley, H.~E.
\newblock (2000) {\em Proc. Natl. Acad. Sci. USA} {\bf 97}, 11149--11152.

\bibitem{watts98}
Watts, D.~J. \& Strogatz, S.~H.
\newblock (1998) {\em Nature} {\bf 393}, 440--442.

\bibitem{havlin00}
Cohen, R., Erez, K., ben Avraham, D.,  \& Havlin, S.
\newblock (2000) {\em Phys. Rev. Lett.} {\bf 85}, 4626--4628.

\bibitem{newman00}
Callaway, D.~S., Newman, M. E.~J., Strogatz, S.~H.,  \& Watts, D.~J.
\newblock (2000) {\em Phys. Rev. Lett.} {\bf 85}, 5468--5471.

\bibitem{barabasi00}
Albert, R., Jeong, H.,  \& Barab{\'a}si, A.-L.
\newblock (2000) {\em Nature} {\bf 406}, 378--382.

\bibitem{pv01a}
Pastor-Satorras, R. \& Vespignani, A.
\newblock (2001) {\em Phys. Rev. Lett.} {\bf 86}, 3200--3203.

\bibitem{weightbook}
Clark, J. \& Holton, D.~A.
\newblock (1998) {\em A first look at graph theory}.
\newblock (World Scientific, Singapore).

\bibitem{yookwt}
Yook, S.~H., Jeong, H., Barab{\'a}si, A.-L.,  \& Tu, Y.
\newblock (2001) {\em Phys. Rev. Lett.} {\bf 86}, 5835--5838.

\bibitem{china}
Li, W. \& Cai, X.
\newblock (2003) Statistical analysis of airport network of china.
\newblock e-print cond-mat/0309236.

\bibitem{luisair}
Guimer{\`a}, R., Mossa, S., Turtschi, A.,  \& Amaral, L. A.~N.
\newblock (2003).
\newblock submitted.

\bibitem{newman01}
Newman, M. E.~J.
\newblock (2001) {\em Phys. Rev. E} {\bf 64}, 016131;
\newblock (2001) {\em Phys. Rev. E} {\bf 64}, 016132.

\bibitem{vicsek01}
Barab\'asi A.~L., Jeong H., N\'eda Z., Ravasz E., Schubert A. \&
Vicsek T.
\newblock (2001) {\em Physica A} {\bf 299}, 559.

\bibitem{freeman77}
Freeman, L.~C.
\newblock (1977) {\em Sociometry} {\bf 40}, 35--41.

\bibitem{goh01}
Goh, K.-I., Kahng, B. \& Kim D. 
\newblock (2001) {\em Phys. Rev. Lett.} {\bf 87}, 278701.

\bibitem{notebet}
More precisely, if ${\cal D}_{hj}$ is the total number of shortest
paths from $h$ to $j$ and ${\cal D}_{hj}(i)$ is the number of these
shortest paths that pass through the vertex $i$, the betweenness of
the vertex $i$ is defined as $b_i=\sum {\cal D}_{hj}(i)/{\cal
D}_{hj}$, where the sum runs over all $h,j$ pairs with $j \neq h\neq
i$. An efficient algorithm to compute betweenness centrality is
reported Brandes, U.
\newblock (2001) {\em Journal of Math. Sociology} {\bf 25}, 163--177.

\bibitem{alexei02}
V{\'a}zquez, A., Pastor-Satorras, R.,  \& Vespignani, A.
\newblock (2002) {\em Phys. Rev. E} {\bf 65}, 066130.

\bibitem{ravasz02}
Ravasz, E. \& Barab{\'a}si, A.-L.
\newblock (2003) {\em Phys. Rev. E} {\bf 67}, 026112.

\bibitem{alexei}
Pastor-Satorras, R., V{\'a}zquez, A.,  \& Vespignani, A.
\newblock (2001) {\em Phys. Rev. Lett.} {\bf 87}, 258701.

\bibitem{maslov02}
Maslov, S. \& Sneppen, K.
\newblock (2001) {\em Science} {\bf 296} 910.

\bibitem{assortative}
Newman, M. E.~J.
\newblock (2002) {\em Phys. Rev. Lett.} {\bf 89}, 208701.

\bibitem{richclub}
Zhou, S. \& Mondragon, R.~J.
\newblock (2003) The missing links in the {BGP}-based {AS} connectivity maps.
\newblock e-print cs.NI/0303028.

\end{thebibliography}
\end{document}